\documentstyle[axodraw,epsf,12pt]{article}
\newcommand{\epem}{e^+e^-}
\begin{document}
\begin{center}
{\Large\bf 
On double $1^{--}$ charmonium production through two-photon $\epem$ annihilation at $\sqrt{s}=10.6\,{\rm GeV}$
}\\
A.V.Luchinsky\footnote{mpti9502@mx.ihep.su}\\
{\it Institute of High Energy Physics, Protvino, Russia}
\end{center}

{\small{\bf Abstract:} Recent measurements of the Belle Collaboration of the exclusive production of two charmonia in the $e^+e^-$ annihilation differ substantially from theoretical predictions. Recently it was suggested that a significant part of this discrepancy can be explained by the process $e^+e^-\to2J/\psi$ and value for its cross section was published. In this paper it is shown, that these results are incorrect and the values for the cross sections of the production of different vector charmonia are presented.
}
\vskip 1cm

The charmonium states (such as $J/\psi$, $\psi(2S)$ or $\eta_c$) are of great interest both from theoretical and experimental points of view because of (i) their clear experimental signature and (ii) great theoretical simplifications which arise from their non-relativistic nature. Especially interesting are reactions of exclusive double charmonium-states production in $e^+e^-$ annihilations since all non-pertubative constants can be determined phenomenologically from $(c\bar c)\to e^+e^-$ decay rates.

Recently the Belle Collaboration has observed $e^+e^-$ annihilation in two charmonium states at center-of-mass energy $\sqrt{s}=10.6 $GeV \cite{Belle} and their measurements differ significantly from the results based on non-relativistic quantum chromodynamics (NRQCD) \cite{NRQCD,Braaten,Liu}. For example the measured cross section for the process $\epem\to J/\psi\eta_c$ is about one order of magnitude higher then the results obtained in \cite{Braaten,Liu,Likh}. This presents a challenge to our current understanding of charmonium production.

The possible way to explain the part of this difference was proposed in  \cite{BBL} (BBL). The authors assumed that some of the events in Belle's $J/\psi\eta_c$ signal could actually be $J/\psi J/\psi$ events produced in two virtual photons $\epem$ annihilation and obtained the values
\begin{eqnarray}
\sigma_{BBL}[\epem\to J/\psi+J/\psi]&=&8.70\pm2.94\,{\rm fb}\label{BBL},\\
\sigma_{BBL}[\epem\to J/\psi+\psi(2S)]&=&7.22\pm2.44\,{\rm fb}\nonumber,\\
\sigma_{BBL}[\epem\to \psi(2S)+\psi(2S)]&=&1.50\pm0.51\,{\rm fb}\nonumber.
\end{eqnarray}
These results have some drawbacks that follow from the approximations used by the authors:

(i) The calculations in BBL were held with the help of NRQCD and instead of physical masses of vector mesons the values $M_{J/\psi}=M_{\psi(2S)}=2m_c=2.8\,{\rm GeV}$ were used. Since in the experiment the reaction $\epem\to VV$ goes near the threshold value of the energy cross section depends strongly on final particle masses and this approximation can lead to large errors.

(ii) In BBL all QCD corrections were neglected. As it was shown in \cite{NRQCD} the effect of these corrections can be significant. For example the leading order value of the probability factor used in \cite{BBL} $[\left<O_1\right>_{J/\psi}]_{LO}=0.208\,{\rm Gev}^3$, but if first order QCD corrections were taken into account it becomes $[\left<O_1\right>_{J/\psi}]_{NLO}=0.335\,{\rm GeV}^3$.

Both of these drawbacks could be corrected if we use not model but physical values of input parameters, i.e. vector meson masses and eeV coupling constants.


\begin{figure}
\begin{picture}(200,100)(0,0)
\ArrowLine(20,80)(80,80)\ArrowLine(80,80)(80,20)\ArrowLine(80,20)(20,20)\Vertex(80,80){1}\Vertex(80,20){1}
\Text(10,80)[]{$e^-$}\Text(10,20)[]{$e^+$}
\Text(50,73)[t]{$k_1$}\Text(75,50)[r]{$q_a$}\Text(50,25)[b]{$k_2$}
\Photon(80,80)(120,80){2}{5}\Photon(80,20)(120,20){2}{5}
\Text(100,75)[t]{$p_1$}\Text(100,25)[b]{$p_2$}
\CArc(125,80)(5,90,270)\ArrowLine(125,85)(175,85)\ArrowLine(175,75)(125,75)
\GOval(175,80)(5,2)(1){0}
\Text(180,80)[l]{$V_1$}
\CArc(125,20)(5,90,270)\ArrowLine(125,15)(175,15)\ArrowLine(175,25)(125,25)
\GOval(175,20)(5,2)(1){0}\Text(180,20)[l]{$V_2$}
\Text(100,0)[b]{(a)}
\end{picture}
\begin{picture}(100,100)(0,0)
\ArrowLine(20,80)(80,80)\ArrowLine(80,80)(80,20)\ArrowLine(80,20)(20,20)\Vertex(80,80){1}\Vertex(80,20){1}
\Text(10,80)[]{$e^-$}\Text(10,20)[]{$e^+$}
\Text(50,73)[t]{$k_1$}\Text(75,50)[r]{$q_b$}\Text(50,25)[b]{$k_2$}
\Photon(80,80)(120,20){2}{5}\Photon(80,20)(120,80){2}{5}
\Text(100,75)[t]{$p_1$}\Text(100,25)[b]{$p_2$}
\CArc(125,80)(5,90,270)\ArrowLine(125,85)(175,85)\ArrowLine(175,75)(125,75)
\GOval(175,80)(5,2)(1){0}
\Text(180,80)[l]{$V_2$}
\CArc(125,20)(5,90,270)\ArrowLine(125,15)(175,15)\ArrowLine(175,25)(125,25)
\GOval(175,20)(5,2)(1){0}\Text(180,20)[l]{$V_1$}
\Text(100,0)[b]{(b)}
\end{picture}

\begin{picture}(200,100)(0,0)
\ArrowLine(20,80)(80,80)\ArrowLine(80,80)(80,20)\ArrowLine(80,20)(20,20)\Vertex(80,80){1}\Vertex(80,20){1}
\Text(10,80)[]{$e^-$}\Text(10,20)[]{$e^+$}
\Photon(80,80)(120,80){2}{5}\Photon(80,20)(120,20){2}{5}
\CArc(125,80)(5,90,270)\ArrowLine(125,85)(175,85)\Line(175,75)(125,25)
\GOval(175,80)(5,2)(1){0}\Text(180,80)[l]{$V_1$}
\CArc(125,20)(5,90,270)\ArrowLine(125,15)(175,15)\Line(175,25)(125,75)
\GOval(175,20)(5,2)(1){0}\Text(180,20)[l]{$V_2$}
\Text(100,0)[b]{(c)}
\end{picture}
\begin{picture}(120,100)(0,0)
\ArrowLine(20,80)(80,80)\ArrowLine(80,80)(80,20)\ArrowLine(80,20)(20,20)\Vertex(80,80){1}\Vertex(80,20){1}
\Text(10,80)[]{$e^-$}\Text(10,20)[]{$e^+$}
\Photon(80,20)(120,80){2}{5}\Photon(80,80)(120,20){2}{5}
\CArc(125,80)(5,90,270)\ArrowLine(125,85)(175,85)\Line(175,75)(125,25)
\GOval(175,80)(5,2)(1){0}\Text(180,80)[l]{$V_2$}
\CArc(125,20)(5,90,270)\ArrowLine(125,15)(175,15)\Line(175,25)(125,75)
\GOval(175,20)(5,2)(1){0}\Text(180,20)[l]{$V_1$}
\Text(100,0)[b]{(d)}
\end{picture}

\caption{QED diagrams for $\epem\to2\gamma^*\to V_1V_2$}\label{fig1}
\end{figure}

Four diagrams of the process $\epem\to V_1V_2$ are shown on figure \ref{fig1}, other two can be obtained from diagrams \ref{fig1}.c and 1.d by interchanging final state vector mesons. First of all, one can note, that diagrams 1.a and 1.b are enhanced with respect to others by the factor $\sim(s/4M_V^2)^2\sim10$, which arises from virtual-photon propagators. The other source of this enhancement is that diagrams 1.a and 1.b are strongly peaked near the beam direction. The QCD corrections caused by the gluon emitted and captured by the same quark or by the quark-antiquark pair which forms a vector meson are taken into account in eeV coupling constant. When gluon line connects quarks of different vector mesons we get additional color and photon-propagator suppressions.
So from now on only diagrams 1.a, 1.b will be considered.

The amplitude for this diagrams is 
\begin{eqnarray*}
{\cal M}&=&e^2g_1g_2\phi_1^\mu\phi_2^\nu\left[
\frac{\bar v(k_2)\gamma_\nu\hat q_a\gamma_\mu u(k_1)}{q_a^2}+
\frac{\bar v(k_2)\gamma_\mu\hat q_b\gamma_\nu u(k_1)}{q_b^2}
\right],
\end{eqnarray*}
where $\phi_i^\alpha$ is i's meson polarization vector, $g_i$ is the  coupling constant of the vertex $eeV_i$  which can be obtained from the electron decay width of the vector meson $V_{1,2}$:
\begin{eqnarray*}
g^2_i&=&12\pi\frac{\Gamma_i^{ee}}{M_i},
\end{eqnarray*}
$M_i$ is the mass of $V_i$ and $\Gamma_i^{ee}$ is the widths of its electron decay.
 
The differential cross section
\begin{eqnarray}
\frac{d\sigma}{dx}&=&\frac{1}{64\pi}\frac{2b}{s^2}\sum|{\cal M}|^2=\frac{g_1^2g_2^2b}{32\pi s^2}\left[
\frac{6s(M_1^2+M_2^2)}{M_1^2M_2^2}-6+\frac{4a^2}{M_1^2M_2^2}-\right.\nonumber\\
&&\left.\frac{3M_1^2}{M_2^2}-\frac{3M_2^2}{M_1^2}-\frac{s^2}{M_1^2M_2^2}+12\frac{a^2+b^2x^2}{a^2-b^2x^2}-
8M_1^2M_2^2\frac{a^2+b^2x^2}{(a^2-b^2x^2)^2}+\right.\nonumber\\
&&2s\left(\frac{M_1^2}{M_2^2}\frac{1}{a-bx}+\frac{M_2^2}{M_1^2}\frac{1}{a+bx}\right)
+2\left(\frac{M_1^2}{M_2^2}\frac{a+bx}{a-bx}+\frac{M_2^2}{M_1^2}\frac{a-bx}{a+bx}\right)+\nonumber\\
&&2\left(\frac{1}{M_1^2}\frac{(a-bx)^2}{a+bx}+\frac{1}{M_2^2}\frac{(a+bx)^2}{a-bx}\right)-
2\left(\frac{a-bx}{M_2^2}+\frac{a+bx}{M_1^2}\right)-\nonumber\\
&&\left.2s^2\left(\frac{1}{M_2^2(a-bx)}+\frac{1}{M_1^2(a+bx)}\right)+
\frac{4sa}{a^2-b^2c^2}\right].\label{ds}
\end{eqnarray}
where $x=\cos\theta$ is the cosine of the angle between $e^-$ and $V_1$ momenta,
\begin{eqnarray*}
a&=&\frac{s-M_1^2-M_2^2}{2},\\
b&=&\frac{1}{2}\sqrt{s-(M_1+M_2)^2}\sqrt{s-(M_1+M_2)^2}.
\end{eqnarray*}

The plot for differential cross section for the case $V_1=V_2=J/\psi$ is shown on figure 2. One can easily see that, as it was mentioned above, it is peaked near the values $x=\pm1$.

The result for total cross section is obtained from (\ref{ds}) by integrating x from -1 to 1:
\begin{eqnarray}
\sigma&=&\int\limits_{-1}^1\frac{d\sigma}{dx}dx=
\frac{g_1^2g_2^2}{8\pi a s^2}\left( (s^2+(M_ 1^2+M_2^2)^2)\log\frac{a+b}{a-b}-8ab\right).\label{s}
\end{eqnarray}
In the case of identical final particles (\ref{s}) should be divided by 2 to avoid double counting.

\begin{table}[t]
\label{tab1}
$$
\begin{array}{|c||c|c|c|c|c|c|}
\hline
i & J/\psi(1S) & \psi(2S) & \psi(3770) & \psi(4040) & \psi(4160) & \psi(4415) \\
\hline
M_i,{\rm GeV} & 3.097 & 3.685 & 3.77 & 4.04 & 4.159 & 4.415 \\
\Gamma^{ee}_i, {\rm eV} & 5.26 & 2.12 & 0.26 & 0.75 & 0.77 & 0.47 \\
\hline
\end{array}$$\caption{}\end{table}

\begin{table}[t]
\label{tab2}
$$
\begin{array}{|c||c|c|c|c|c|c|}
\hline
& J/\psi(1S) & \psi(2S) & \psi(3770) & \psi(4040) & \psi(4160) & \psi(4415) \\
\hline\hline
J/\psi(1S) & 2.26 & 1.46 & 0.17 & 0.46 & 0.46 & 0.26 \\
\psi(2S)   & 1.46 & 0.23 & 0.06 & 0.15 & 0.15 & 0.08 \\
\psi(3770) & 0.17 & 0.06 & 0.003 & 0.02 & 0.02 & 0.01 \\
\psi(4040) & 0.46  & 0.15 & 0.02  & 0.02 & 0.05 & 0.03 \\
\psi(4160  & 0.46  & 0.15 & 0.02 & 0.05  & 0.02 & 0.02 \\
\psi(4415) & 0.26 & 0.08 & 0.01 & 0.03 & 0.02 & 0.01 \\
\hline
\end{array}
$$
\caption{$\sigma(\epem\to V_1V_2)$, fb}
\end{table}

Using experimental values of vector meson masses and electron decay widths listed in table 1 one can obtain total cross-sections presented in table 2. These results differ significantly from those presented in BBL, for example for $V_1=V_2=J/\psi$
\begin{eqnarray*}
\frac{\sigma_{BBL}(\epem\to J/\psi J/\psi)}{\sigma(\epem\to J/\psi J/\psi)}\approx3.8.
\end{eqnarray*}

Thus the author would like to note that the explanation proposed in BBL is unsatisfactory and the question of the difference between results of the Belle experiment and their theoretical predictions remains opened.

Author would like to thank A.K.Likhoded and S.S.Gershrein for valuable discussions.

\begin{figure}
\begin{picture}(100,200)
\put(10,10){\epsfxsize=14cm \epsfbox{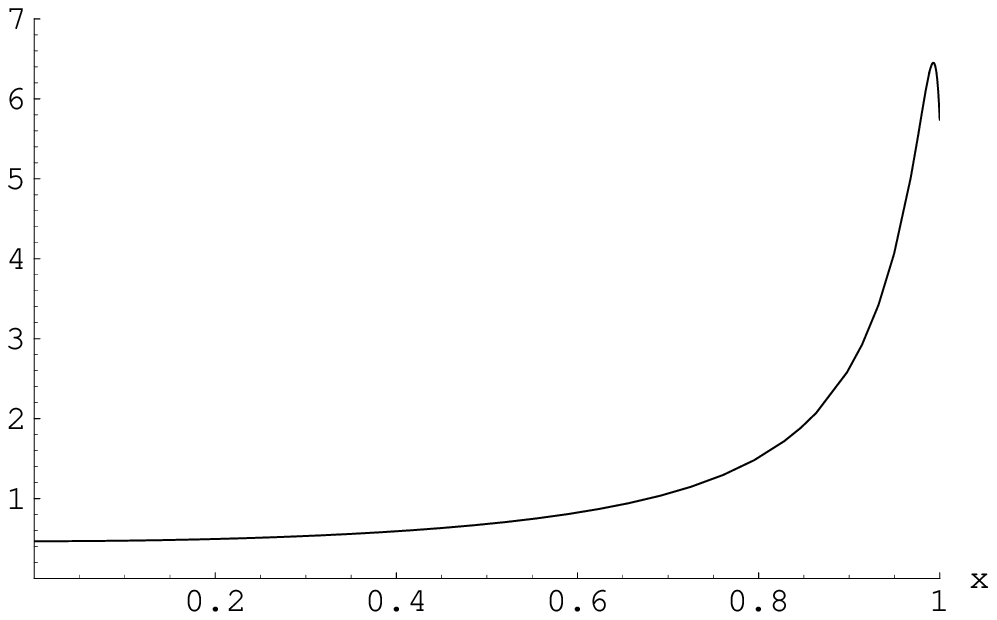}}
\put(30,250){$\frac{d\sigma(\epem\to2J/\psi)}{dx}\,,{\rm fb}$}
\end{picture}
\caption{Angular distribution for the process $\epem\to2J/\psi$}
\end{figure}



\begin{thebibliography}{99}
\bibitem{Belle}
K.Abe { et al.}, [BELLE Collaboration], Phys.Rev.Lett. {\bf89}, 142001 (2002)
\bibitem{NRQCD}
G.T. Bodwin, E.Braaten and G.P.Lepage, Phys. Rev. D{\bf51}, 1125 (1995); {\bf55} 5853(E); hep-ph/9407339
\bibitem{Braaten}
E.Braaten and J.Lee, hep-ph/0211085
\bibitem{Liu}
K.-Y.Liu, Z.-G.He and K.-T.Chao, hep-ph/0211181
\bibitem{Likh}
A.V. Berezhnoy, A.K.Likhoded, private communication
\bibitem{BBL}
G.T.Bodwin, E.Braaten, J.Lee, hep-ph/0212181
\end{thebibliography}
\end{document}